\documentclass{article}
\usepackage[utf8]{inputenc}
\usepackage{graphicx}
\usepackage{hyperref}
\usepackage{color}
\usepackage [english]{babel}
\usepackage [autostyle, english = american]{csquotes}
\MakeOuterQuote{"}
\begin{document}

\begin{titlepage}
	\centering
	
	{\scshape\huge Is Smaller Better: A Proposal To Consider Bacteria For Biologically Inspired Modeling\par}
	\vspace{5.5cm}
	\vspace{1cm}
	{\Large Archana Ram*$^{1}$ and Andrew Lo$^{1,2}$\par}
	\vfill

	{$^1$ Department of Electrical Engineering and Computer Science, Massachusetts Institute of Technology, Cambridge, Massachusetts 02139, $^2$ Sloan School of Management, Massachusetts Institute of Technology, Cambridge, Massachusetts 02139
	\\\textit{*Email: archanar@mit.edu}}
\end{titlepage}

\section*{\textit{Abstract}}

Bacteria are easily characterizable model organisms with an impressively complicated set of capabilities. Among their capabilities is quorum sensing, a detailed cell-cell signaling system that may have a common origin with eukaryotic cell-cell signaling. Not only are the two phenomena similar, but quorum sensing, as is the case with any bacterial phenomenon when compared to eukaryotes, is also easier to study in depth than eukaryotic cell-cell signaling. This ease of study is a contrast to the only partially understood cellular dynamics of neurons. Here we review the literature on the strikingly neuron-like qualities of bacterial colonies and biofilms, including ion-based and hormonal signaling, and action potential-like behavior. This allows them to feasibly act as an analog for neurons that could produce more detailed and more accurate biologically-based computational models. Using bacteria as the basis for biologically feasible computational models may allow models to better harness the tremendous ability of biological organisms to make decisions and process information. Additionally, principles gleaned from bacterial function have the potential to influence computational efforts divorced from biology, just as neuronal function has in the abstract influenced countless machine learning efforts.

\section*{Introduction}

The number of bacteria on Earth is staggering. Conservative estimates claim that there are nearly half a million bacterial species in just 30 grams of soil.$^{43}$ The myth persists that bacteria are simple organisms, but this could not be further from the truth. The complexity of bacterial function in many ways mirrors that of eukaryotic cells.$^{19}$ In this review, we examine the literature demonstrating that bacterial cells, colonies, and biofilms exhibit notable similarities to neurons and neuronal networks, including action potential-like behavior, ion-based signaling, and hormonal signaling.
\\
\\
A natural first question is "why would one even consider using bacteria instead of neurons?" This can be answered in a few ways. First, bacteria are simple. Take two well-studied model bacteria, \textit{Escherichia coli} and \textit{Bacillus subtilis}. Their genomes are both slightly over 4Mbp long and code roughly 4000 protein-coding genes.$^{59}$ Compare this with the human genome, which is roughly 3 billion base pairs long and codes for roughly 100,000 proteins. Even \textit{C. elegans}, one of the simplest neuroscientific model organisms, has a genome size of 100Mbp [25x the size of a bacterial genome] and codes for almost 22,000 proteins.$^{61}$ The comparative simplicity of bacteria makes them inherently an easier organism to study. There is, simply put, less they are capable of and, as a result, there is less to understand about their functionality. Additionally, they are easier to work with in a laboratory environment. One \textit{E. coli} cell divides roughly every 20-60 minutes and, as a result, colonies can be grown overnight. This, coupled with the fact that bacteria are able to become competent and uptake extracellular DNA, allows for easier and quicker genetic experimentation.$^{58}$ Neurons, on the other hand, are terminally differentiated cells. As a result, division is much slower.$^{60}$  
\\
\\
One major benefit of the use of bacteria as a model organism is the fact that each bacterium is an organism to itself as opposed to a cell constituting part of a larger organism. Take the phenomenon of bacterial chemotaxis, which can easily be investigated in a laboratory setting. A chemical attractant can be placed into either a suspension or a plate of bacteria and each individual organism will move towards the attractant.$^{18}$ This sort of well-characterized input-output behavior cannot be replicated in neuronal cell cultures and, as a result, allows for more complex behavioral experimentation that may actually help elucidate comparable functionality in higher-order organisms, as will be discussed in the next section.  
\\
\\
This paper does not necessarily seek to propose the use of bacteria as a model organism for purely behavioral work but rather to help create more biologically feasible models. In order to create a biologically plausible model, one must first possess a thorough understanding of the biological underpinnings of the model, something that cannot be said to be the case with regards to the mammalian brain or even lower-order nervous systems owing to their tremendous complexity. Despite the general lack of understanding of neuronal network function, artificial neural networks use as their basic unit a binary input/output node that is supposedly an abstraction of a neuron.$^{38}$ This sort of abstraction is an oversimplified view of actual neuronal function that makes the network itself biologically implausible.$^{39}$ This review begins with an overview of similarities between mammals and lower-order organisms and proceeds to discuss the ways in which bacterial communities mirror neuronal circuits and networks. It seeks to propose the investigation of prokaryotes, e.g. \textit{B. subtilis}, in order to gain insight into understanding and better modeling higher-level organisms such as mammals. 

\section*{Similarities between bacteria and higher organisms}

Continuing with the earlier discussion about bacterial chemotaxis, hunger and satiety detection is a relatively well-conserved system among many different species and, as a result, is a good starting point for a discussion about inter-species similarities.$^{44, 45, 46, 47, 48}$ There exist numerous structural and molecular similarities between \textit{Drosophila melanogaster} and mammals, for example, in this regard, thereby allowing \textit{Drosophila} to potentially serve as a easier-to-study proxy for the mammalian brain in this context.$^{49, 50, 51, 52, 53, 54, 55, 56, 57}$ These similarities are also seen with respect to eukaryotes such as \textit{Drosophila} and lower-level organisms. When hungry, prokaryotes and \textit{Drosophila} larvae ascend nutrient gradients in a process known as chemotaxis. \textit{Drosophila} larvae's approach is in stages--first, approach the source then, once near the source, reach it, overshoot it and then return to the source. Their motion consists of "runs" and "turns"--the "runs" predominate their motion and the turns are abrupt--generally when a decreasing chemical concentration is sensed during forward motion. The sort of motion is akin to a biased random walk in that the organism aims to meander towards the center of the nutrition concentration but may wander slightly along the way.$^{16, 17, 18}$ This is similar to the mechanism employed by \textit{E. coli}, which also seem to also favor crawls towards higher nutrient concentrations as opposed to lower ones.$^{18}$
\\
\\
The similarities between \textit{Drosophila} and bacteria go further than feeding behaviors. Consider quorum sensing, a form of bacterial cell-cell communication used to sense local bacterial population density. The protein AarA of the Gram-negative soil bacterium \textit{Providencia stuartii} is necessary to release the molecular signals for quorum sensing in that species. This protein, however, appears to be homologous to the \textit{Drosophila} protease RHO, which is required to activate epidermal growth factor receptor ligands in that organism, in addition to being essential to its proper wing vein development and eye organization. Indeed, the two are so similar chemically that expressing RHO in \textit{P. stuartii} AarA acts as a substitute for AarA expression, as the mutants possess relatively normal quorum sensing capabilities. Similarly, expressing \textit{P. stuartii} AarA in \textit{Drosophila} RHO mutants allows wing development to proceed normally, again allowing the substitution of the two homologs, despite their origin in two very different species.$^{26}$
\\
\\
This homology is not an isolated incident. It appears that many signaling mechanisms are shared by prokaryotes and eukaryotes. In fact, the evolution of cell-cell signaling is hypothesized to have been more reliant on horizontal gene transfer from bacteria to animals than purely vertical inheritance.$^{32}$ An interesting example is glutamate decarboxylase, which catalyzes the amino acid glutamate to form the neurotransmitter GABA. This enzyme is coded by a gene acquired by eukaryotes from prokaryotes through horizontal gene transfer.$^{26}$

\section*{Similarities in bacterial and neuron ion-based communication}

Bacteria not only have influenced the development of eukaryotic cell-cell signaling, they possess a number of direct similarities to neurons, specifically in their means of cell-cell communication and the cell membrane.
\\
\\
The neuronal membrane voltage is regulated by the common but important ions Na$^+$, Cl$^-$, Ca$^{2+}$ and K$^+$. Briefly, K$^+$ tends to accumulate inside the membrane, while Na$^+$, Cl$^-$ and Ca$^{2+}$ have higher concentrations outside the membrane. Notably, K$^+$ is a major influence on membrane voltage.$^{27}$ There is a growing body of evidence that bacteria also use these ions to regulate voltage across the bacterial cell membrane. As in neurons, Na$^+$ accumulates on the outside of the membrane, while K$^+$ accumulates inside. There is even evidence for ionic Na$^+$/K$^+$ exchange, perhaps mediated by pumps similar to the ones found in neurons.$^{28}$ To this end, it should be pointed out that the first structure of the K$^{+}$ channel, essential to the function of both bacteria and neurons, was first determined from a bacterial source due to the ease of bacterial study compared to neurons.$^{36}$ The resting membrane potential of \textit{E. coli} is -75 mV, only about 5 mV lower than that of neurons, suggesting additional electrophysiological parallels between the two cell types.$^{29}$ 
\\
\\
These ions are used in bacteria not only to regulate membrane voltage, but also as signals. The PhoP/PhoQ system in \textit{Salmonella typhimurium} governs various virulence properties of that organism, and has distinct binding sites for both Ca$^{2+}$ and Mg$^{2+}$. These extracellular ions act as the signals that instigate the action of the system in a way that appears somewhat analogous to the neuronal calcium channel regulator CaBP1, which also has binding sites for both Ca$^{2+}$ and Mg$^{2+}$.$^{30, 31, 63}$
\\
\\
Another example of ion-based signaling is seen in \textit{Bacillus subtilis}, a Gram-positive spore-forming bacterium.$^{21}$ Bacteria tend to produce biofilms when they are stressed, e.g. when there are limited nutrients in the environment.$^{34}$ When the species produces biofilms of greater than 1 million cells, the colony naturally produces electrical oscillations that serve to modulate the biofilm's voltage as a whole. Intracellular and extracellular potassium ions produce a gradient on the given substrate, towards which motile bacteria of various species are attracted, based on the potassium's capability to alter their resting membrane potential. This potassium-based attraction appears to be coupled to the biofilm's oscillations, thereby producing a phenomenon reminiscent of graded action potential-based neuronal signaling in higher organisms [Figure 1].$^{22}$ 

\begin{figure}
    \centering
    \includegraphics[scale=0.8]{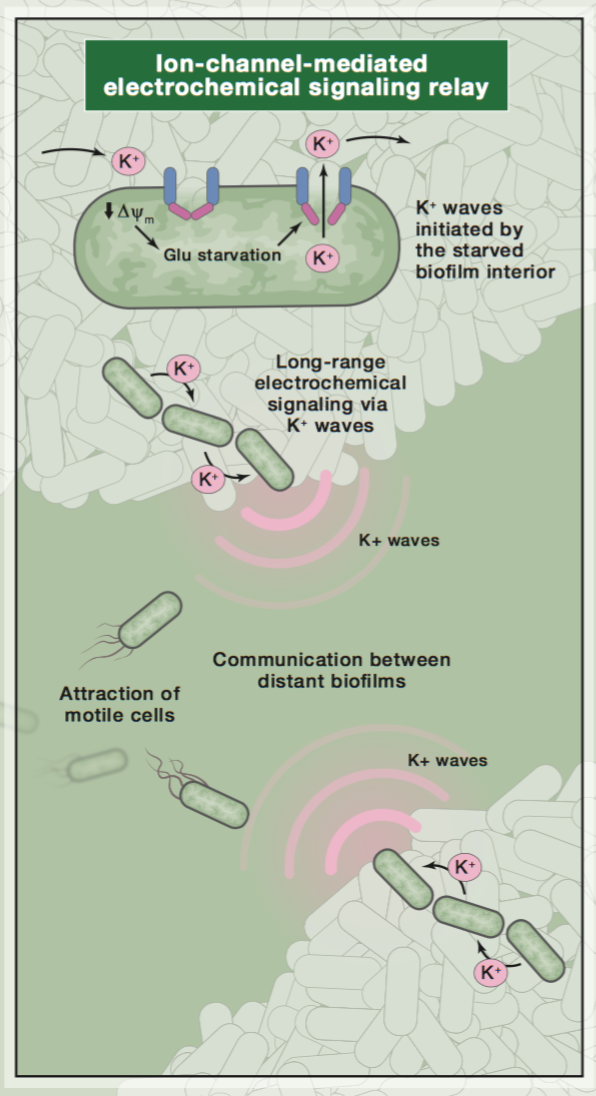}
    \caption{A diagram of ion-based communication in biofilms. Image reused with permission from original publisher.$^{22}$}
\end{figure}

\section*{Similarities between bacterial quorum sensing and neuronal communication}

Ions, however, are not the only sort of signals present in bacteria. Similar to neurons, bacteria also use hormonal compounds to communicate. Bacterial quorum sensing is achieved through the use of different peptides and hormones, and in many ways, it mimics how neurons communicate. There are several types of quorum sensing systems, classified by signaling molecule. One system involves hormone-like compounds known as autoinducers (AIs), found in Gram-negative bacteria. Frequently used and studied among these autoinducer compounds is the amino acid derivative AHL (\textit{N}-acyl homoserine lactone). AHLs are mainly used for intra-species communication between Gram-negative bacteria, useful in environments where different bacterial species share resources. The most common autoinducer used in Gram-negative bacteria, mainly used for communication within a given colony, is AI-2. This is a relatively universal communication molecule used in over 40 species of bacteria. A divergent type of quorum sensing system is found in Gram-positive bacteria. Rather than autoinducer
molecules, it uses modified oligopeptides to signal population
density.$^{33}$
\\
\\
The canonical example of quorum sensing is the system used by \textit{Vibrio fischeri}, a bacterium that lives inside the light organ of the squid \textit{Euprymna scolopes}. Once these bacteria grow to a high enough cell density, quorum sensing allows them to induce the expression of genes necessary for bioluminescence.$^{26}$ At its most basic, quorum sensing molecules passively diffuse through the bacterial membrane, accumulating both intra- and extracellularly to a concentration proportional to cell density. Once the signal has reached an appropriate level inside the cell, the transcription of certain genes will begin. This signal, however, can also be detected through receptors. Gram-negative/AI2-based systems tend to use cytoplasmic receptors but, interestingly, gram-positive bacteria like \textit{B. subtilis} that exhibit oscillatory electrical communication similar to neurons also possess membrane receptors similar to those in neurons.$^{62}$ Quorum sensing can occur in series and in parallel. It can induce the transient expression of genes, and it can be used by one bacterial colony to "eavesdrop" on other populations. There even exist hierarchical quorum sensing circuits. Sometimes quorum sensing will produce inhibitory signals within a colony, while in other cases, bacterial populations will "sabotage" a quorum sensing signal from another colony, and degrade it in a process known as quorum quenching. \textit{B. subtilis}, for example, produces an enzyme called AiiA that is capable of hydrolyzing the AHL of another soil bacterium, thereby inhibiting its external signaling attempts.$^{26, 33}$
\\
\\
There exist important and striking similarities between quorum sensing and neuronal communication. Both quorum sensing and neuronal circuits operate in series and in parallel, and both are able to develop into a hierarchical multi-circuit system.$^{35}$ Neurons can provide both excitatory and inhibitory signals, just as quorum sensing does, and neurons can communicate with each other as well as other cell types, just as quorum sensing can function between bacteria of the same species or between different species.$^{27}$ These parallels are unmistakable; it is easy to confirm that there are at least abstract similarities between bacterial colony and neuronal function.

\section*{Abstracted bacterial systems are similar to neuronal analogues}

Perhaps the most compelling evidence supporting this analogy, especially from a computational perspective, comes from the bacterial attractant concentration detection system. When unsaturated, this system can be approximated and therefore modeled as a linear time-invariant (LTI) system whose impulse response is depicted below [Figure 2]. This response curve is almost identical to an action potential, which one might consider to be the "impulse response" of an individual neuron receiving external stimulation.$^{41}$
\\
\begin{figure}
    \centering
    \includegraphics[width=\textwidth]{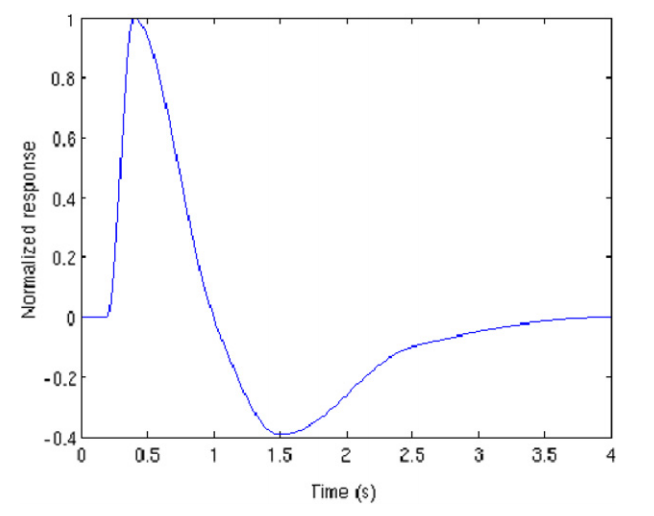}
    \includegraphics[width=\textwidth]{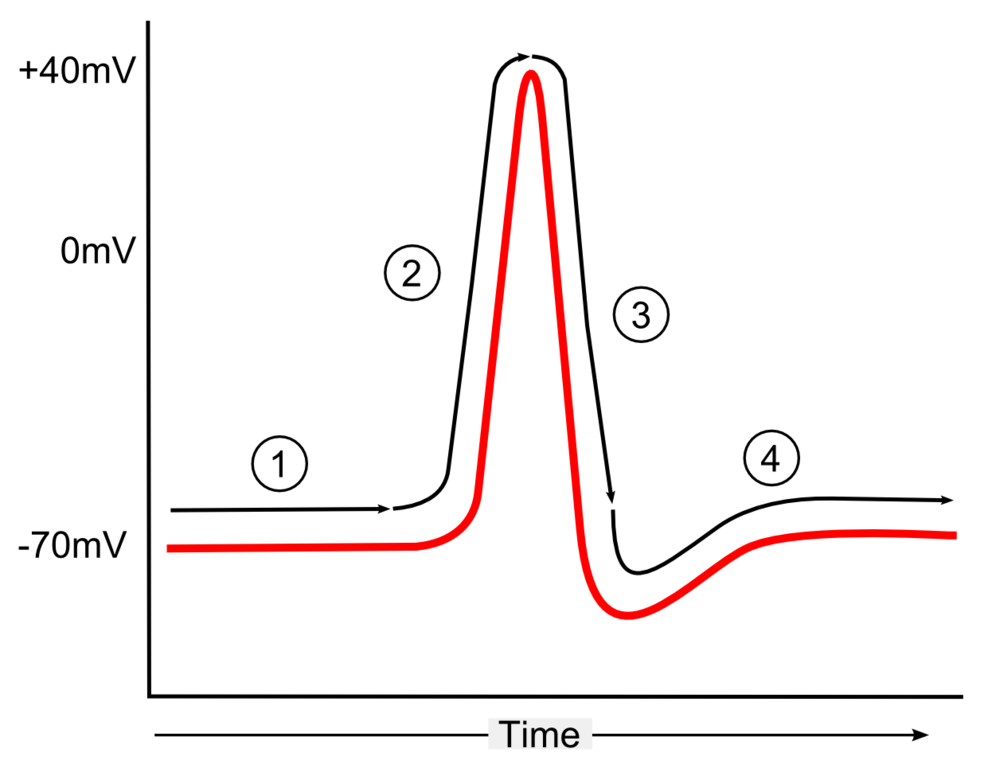}
    \caption{Above, the impulse response for the LTI approximation of a bacterial concentration-detection system, reprinted with permission of the authors.$^{41}$ Below, a characteristic action potential graph.$^{42}$}
\end{figure}
\\
This result demonstrates how modeled bacterial behavior can foreseeably serve as a realistic proxy to modeled neuronal behavior. Bacteria have many benefits over neurons in a computational sense--most importantly, they are functionally simpler than neurons and easier to observe in a non-computational biological context. The comparatively easier-to-study nature of a bacterial population, combined with its complex neuron-like behavior, makes the bacterium an interesting and unconventional candidate for neuroscience research. The relative simplicity of bacteria allows for ease of biologically faithful modeling, while the complexity of their cell-cell interactions has the potential to yield insights into higher-level behaviors from feeding strategies to distributed processing.

\section*{Conclusion}

The use of bacteria in neuroscience research naturally prompts the question of why one would use organisms without neurons for neuroscience. The use of bacteria in cellular neuroscience may be farfetched, but we believe there is sufficient evidence to merit further study of the similarities between bacteria and neurons. It is important to note, however, that beyond neuroscience, there exist numerous benefits to using bacteria instead of neurons as the basis for biologically inspired computational models.
\\
\\
Consider the artificial neural network. An ANN abstracts neurons to weighted binary input/output units.$^{38}$ It is readily apparent that this abstraction is not only an oversimplification of actual neuronal function, but is also an incorrect representation of actual neuronal activity. The brains of higher organisms (for example, mammals) are extraordinarily complicated. It would be a nontrivial task at best to represent even a sliver of their activity. It is highly unlikely that ANNs in their current state are in any way biologically faithful models. Rather, they are a computational abstraction that has become notably divorced from the reality upon which they are based.$^{39}$
\\
\\
In contrast, consider the bacterium. While it cannot be abstracted to a weighted binary input/output unit, its abilities can be far more easily viewed and understood by the biologist than the neuron. It exhibits clear ion channel-based communication, while bacterial quorum sensing is notably similar to neuronal communication. In addition, when bacteria coalesce into a biofilm, they are able to produce something akin to the graded action potential activity seen in \textit{C. elegans}, the focus of much neuroscientific research over the years.$^{25}$ Again, it is important to emphasize that bacterial behavior is not fundamentally dissimilar from that of higher organisms. This allows us to claim that insights gained from bacterial modeling could apply to similar functions in higher organisms such as \textit{Drosophila} or mammals.
\\
\\
Bacteria are by no means the "perfect" neuroscientific model organism---they lack neurons, to begin with---but they represent a compromise. They exhibit network-like activity, and as a colony, they are capable of making decisions in ways similar to primitive lower organisms.$^{40}$ Their beauty comes in the simplicity and relative ease of study compared to other organisms. For a model to have a sound biological basis, the biology upon which it is based must be well understood. It is intuitive enough to say that the simpler the organism, the easier it is to understand it. Following this logic to its conclusion, bacteria---more specifically, bacterial colonies---are a possible stepping stone for biologically faithful neuroscientific modeling. They have the potential to lead the way to modeling more complex organisms, especially after dedicated behavioral, molecular, and cellular work makes their neuron-like activity less of a black box and more of a white box.
\\
\\
The use of modeled bacteria has the potential to go far beyond neuroscience, however. One should consider the possible benefits to computer science. By using a modeled bacterium instead of a weighted binary input/output unit, as does an ANN, one may allow for greater network functionality using fewer nodes and a smaller amount of training data. While one should not expect any sort of network based on modeled bacteria to mirror bacterial networks functionally, it is not inconceivable that the additional complexity could enhance network functionality in ways that simple weighted binary input/output units might not.
\\
\\
Finally, why should we bother with more feasible biological models at all? Are the models we have in use today---Bayesian networks, artificial neural networks, and so forth---not enough for our needs? The answer becomes obvious if we rephrase the question: Is it worth it to allow the insights gained from biological work to influence computational work? It does not go without saying that the mammalian brain is an incredibly compact, tremendously powerful organ which the most powerful models and computers cannot match. It would be foolish not to work towards a time when the mammalian brain is not only understandable, but influential in the creation of both software and hardware. Modeling bacterial colonies is a first step in helping ensure this possibility may one day become a reality.

\section*{Acknowledgements}
This publication would not have been possible without the dedicated guidance of Jessie Stickgold-Sarah, Ron Weiss, Patrick Winston, Alan Grossman and Michaela Ennis. The authors would also like to thank Gurol Suel, Guoping Feng, Anil Sindhwani, Wei Low, Vineel Adusumilli, Vipul Vachharajani, Robby Vasen, Ramu Thiagarajan, and Nikhil Kunapuli for their time and help in editing/conceiving of this publication.

\clearpage
\section*{References}

$^1$ Krashes, M. J., DasGupta, S., Vreede, A., White, B., Armstrong, J. D., \& Waddell, S. (2009). A neural circuit mechanism integrating motivational state with memory expression in Drosophila. Cell, 139(2), 416-427.
\\
\\
$^2$ Stuber, G. D., \& Wise, R. A. (2016). Lateral hypothalamic circuits for feeding and reward. Nature neuroscience, 19(2), 198-205. Chicago
\\
\\
$^3$ Tessmar-Raible, K., Raible, F., Christodoulou, F., Guy, K., Rembold, M., Hausen, H., \& Arendt, D. (2007). Conserved sensory-neurosecretory cell types in annelid and fish forebrain: insights into hypothalamus evolution. Cell, 129(7), 1389-1400.
\\
\\
$^4$ Bouret, S. G., \& Simerly, R. B. (2006). Developmental programming of hypothalamic feeding circuits. Clinical genetics, 70(4), 295-301.
\\
\\
$^5$ Jobst, E. E., Enriori, P. J., \& Cowley, M. A. (2004). The
electrophysiology of feeding circuits. Trends in Endocrinology \&
Metabolism, 15(10), 488-499.
\\
\\
$^6$ Denis, R. G., Joly-Amado, A., Webber, E., Langlet, F., Schaeffer, M., Padilla, S. L., ... \& Martinez, S. (2015). Palatability can drive feeding independent of AgRP neurons. Cell metabolism, 22(4), 646-657.
\\
\\
$^7$ Dietrich, M. O., Zimmer, M. R., Bober, J., \& Horvath, T. L. (2015). Hypothalamic Agrp neurons drive stereotypic behaviors beyond feeding. Cell, 160(6), 1222-1232.
\\
\\
$^8$ Krashes, M. J., Koda, S., Ye, C., Rogan, S. C., Adams, A. C., Cusher, D. S., ... \& Lowell, B. B. (2011). Rapid, reversible activation of AgRP neurons drives feeding behavior in mice. The Journal of clinical investigation, 121(4), 1424.
\\
\\
$^9$ Bouret, S. G., Draper, S. J., \& Simerly, R. B. (2004). Trophic action of leptin on hypothalamic neurons that regulate feeding. Science, 304(5667), 108-110.
\\
\\
$^{10}$Stuber, G. D., \& Wise, R. A. (2016). Lateral hypothalamic circuits for feeding and reward. Nature neuroscience, 19(2), 198-205. Chicago
\\
\\
$^{11}$ Alhadeff, A. L., Hayes, M. R., \& Grill, H. J. (2014). Leptin receptor signaling in the lateral parabrachial nucleus contributes to the control of food intake. American Journal of Physiology-Regulatory, Integrative and Comparative Physiology, 307(11), R1338-R1344.
\\
\\
$^{12}$ Olsen, S. R., \& Wilson, R. I. (2008). Cracking neural circuits in a tiny brain: new approaches for understanding the neural circuitry of Drosophila. Trends in neurosciences, 31(10), 512-520.
Chicago
\\
\\
$^{13}$ De Velasco, B., Erclik, T., Shy, D., Sclafani, J., Lipshitz, H., McInnes, R., \& Hartenstein, V. (2007). Specification and development of the pars intercerebralis and pars lateralis, neuroendocrine command centers in the Drosophila brain. Developmental biology, 302(1), 309-323.
\\
\\
$^{14}$ Pool, A. H., Kvello, P., Mann, K., Cheung, S. K., Gordon, M. D., Wang, L., \& Scott, K. (2014). Four GABAergic interneurons impose feeding restraint in Drosophila. Neuron, 83(1), 164-177.
\\
\\
$^{15}$ Chiang, A. S., Lin, C. Y., Chuang, C. C., Chang, H. M., Hsieh, C. H., Yeh, C. W., ... \& Wu, C. C. (2011). Three-dimensional reconstruction of brain-wide wiring networks in Drosophila at single-cell resolution. Current Biology, 21(1), 1-11.
\\
\\
$^{16}$ Gomez-Marin, A., Stephens, G. J., \& Louis, M. (2011). Active sampling and decision making in Drosophila chemotaxis. Nature communications, 2, 441.
\\
\\
$^{17}$ Bargmann, C. I., \& Horvitz, H. R. (1991). Chemosensory neurons with overlapping functions direct chemotaxis to multiple chemicals in C. elegans. Neuron, 7(5), 729-742.
\\
\\
$^{18}$ Berg, H. C., \& Brown, D. A. (1972). Chemotaxis in Escherichia coli analysed by three-dimensional tracking. Nature, 239(5374), 500-504.
\\
\\
$^{19}$ Shapiro, J. A. (1988). Bacteria as multicellular organisms. Scientific American, 258(6), 82-89.
\\
\\
$^{20}$ Shapiro, J. A. (1998). Thinking about bacterial populations as multicellular organisms. Annual Reviews in Microbiology, 52(1), 81-104.
\\
\\
$^{21}$ Aguilar, C., Vlamakis, H., Losick, R., \& Kolter, R. (2007). Thinking about Bacillus subtilis as a multicellular organism. Current opinion in microbiology, 10(6), 638-643.
\\
\\
$^{22}$ Humphries, J., Xiong, L., Liu, J., Prindle, A., Yuan, F., Arjes,
H. A., ... \& S\"uel, G. M. (2017). Species-independent attraction to
biofilms through electrical signaling. Cell, 168(1), 200-209.
\\
\\
$^{23}$ Keller, E. F., \& Segel, L. A. (1971). Model for chemotaxis. Journal of theoretical biology, 30(2), 225-234.
\\
\\
$^{24}$ Hillen, T., \& Painter, K. J. (2009). A user's guide to PDE models
for chemotaxis. Journal of mathematical biology, 58(1-2), 183.
\\
\\
$^{25}$ Lockery, S. R., \& Goodman, M. B. (2009). The quest for action potentials in C. elegans neurons hits a plateau. Nature neuroscience, 12(4), 377-378.
\\
\\
$^{26}$ Waters, C. M., \& Bassler, B. L. (2005). Quorum sensing: cell-to-cell communication in bacteria. Annu. Rev. Cell Dev. Biol., 21, 319-346.
\\
\\
$^{27}$ Bear, M. F., Connors, B. W., \& Paradiso, M. A. (Eds.). (2007).
Neuroscience (Vol. 2). Lippincott Williams \& Wilkins.
\\
\\
$^{28}$ Lanyi, J. K. (1979). The role of Na+ in transport processes of bacterial membranes. Biochimica et Biophysica Acta (BBA)-Reviews on Biomembranes, 559(4), 377-397.
\\
\\
$^{29}$ Schuldiner, S., \& Kaback, H. R. (1975). Mechanisms of active transport in isolated bacterial membrane vesicles. 28. Membrane potential and active transport in membrane vesicles from Escherichia coli. Biochemistry, 14(25), 5451-5461.
\\
\\
$^{30}$ VÃ©scovi, E. G., Ayala, Y. M., Di Cera, E., \& Groisman, E. A. (1997). Characterization of the bacterial sensor protein phoq evidence for distinct binding sites for mg2+ and ca2+. Journal of Biological Chemistry, 272(3), 1440-1443.
\\
\\
$^{31}$ VÃ©scovi, E. G., Soncini, F. C., \& Groisman, E. A. (1996). Mg 2+ as an extracellular signal: environmental regulation of Salmonella virulence. Cell, 84(1), 165-174.
\\
\\
$^{32}$ Hughes, D. T., \& Sperandio, V. (2008). Inter-kingdom signalling: communication between bacteria and their hosts. Nature reviews. Microbiology, 6(2), 111.
\\
\\
$^{33}$ Bassler, B. L. (2002). Small talk: cell-to-cell communication in bacteria. Cell, 109(4), 421-424.
\\
\\
$^{34}$ Prindle, A., Liu, J., Asally, M., Ly, S., Garcia-Ojalvo, J., \&
S\"uel, G. M. (2015). Ion channels enable electrical communication in
bacterial communities. Nature, 527(7576), 59-63.
\\
\\
$^{35}$ Krashes, M. J., DasGupta, S., Vreede, A., White, B., Armstrong, J. D., \& Waddell, S. (2009). A neural circuit mechanism integrating motivational state with memory expression in Drosophila. Cell, 139(2), 416-427.
\\
\\
$^{36}$ Yellen, G. (1999). The bacterial K+ channel structure and its implications for neuronal channels. Current opinion in neurobiology, 9(3), 267-273.
\\
\\
$^{37}$ Roth, G., \& Dicke, U. (2005). Evolution of the brain and
intelligence. Trends in cognitive sciences, 9(5), 250-257.
\\
\\
$^{38}$ Lippmann, R. (1987). An introduction to computing with neural nets. IEEE Assp magazine, 4(2), 4-22.
\\
\\
$^{39}$ Staelin, D. H., \& Staelin, C. H. (2011). Models for Neural Spike
Computation and Cognition. CreateSpace.[Links].
\\
\\
$^{40}$ Ben-Jacob, E., Lu, M., Schultz, D., \& Onuchic, J. N. (2014). The physics of bacterial decision making. Frontiers in cellular and infection microbiology, 4.
\\
\\
$^{41}$ Cobo, L. C., \& Akyildiz, I. F. (2010). Bacteria-based
communication in nanonetworks. Nano Communication Networks, 1(4), 244-256.
\\
\\
$^{42}$ Image sourced from \href{http://hyperphysics.phy-astr.gsu.edu/hbase/Biology/imgbio/actpot.gif}{\textcolor{blue}{https://upload.wikimedia.org/wikipedia/commons/thumb/e/e0/Action\_potential\_schematic.png/984px-Action\_potential\_schematic.png}}
\\
\\
$^{43}$ Dykhuizen, D. E. (1998). Santa Rosalia revisited: why are there so many species of bacteria?. Antonie van Leeuwenhoek, 73(1), 25-33.
\\
\\
$^{44}$ Krashes, M. J., DasGupta, S., Vreede, A., White, B., Armstrong, J. D., \& Waddell, S. (2009). A neural circuit mechanism integrating motivational state with memory expression in Drosophila. Cell, 139(2), 416-427.
\\
\\
$^{45}$ Stuber, G. D., \& Wise, R. A. (2016). Lateral hypothalamic circuits for feeding and reward. Nature neuroscience, 19(2), 198-205. Chicago
\\
\\
$^{46}$ Tessmar-Raible, K., Raible, F., Christodoulou, F., Guy, K., Rembold, M., Hausen, H., \& Arendt, D. (2007). Conserved sensory-neurosecretory cell types in annelid and fish forebrain: insights into hypothalamus evolution. Cell, 129(7), 1389-1400.
\\
\\
$^{47}$ Bouret, S. G., \& Simerly, R. B. (2006). Developmental programming of hypothalamic feeding circuits. Clinical genetics, 70(4), 295-301.
\\
\\
$^{48}$ Jobst, E. E., Enriori, P. J., \& Cowley, M. A. (2004). The
electrophysiology of feeding circuits. Trends in Endocrinology \&
Metabolism, 15(10), 488-499.
\\
\\
$^{49}$ Denis, R. G., Joly-Amado, A., Webber, E., Langlet, F., Schaeffer, M., Padilla, S. L., ... \& Martinez, S. (2015). Palatability can drive feeding independent of AgRP neurons. Cell metabolism, 22(4), 646-657.
\\
\\
$^{50}$ Dietrich, M. O., Zimmer, M. R., Bober, J., \& Horvath, T. L. (2015). Hypothalamic Agrp neurons drive stereotypic behaviors beyond feeding. Cell, 160(6), 1222-1232.
\\
\\
$^{51}$ Krashes, M. J., Koda, S., Ye, C., Rogan, S. C., Adams, A. C., Cusher, D. S., ... \& Lowell, B. B. (2011). Rapid, reversible activation of AgRP neurons drives feeding behavior in mice. The Journal of clinical investigation, 121(4), 1424.
\\
\\
$^{52}$ Bouret, S. G., Draper, S. J., \& Simerly, R. B. (2004). Trophic action of leptin on hypothalamic neurons that regulate feeding. Science, 304(5667), 108-110.
\\
\\
$^{53}$Stuber, G. D., \& Wise, R. A. (2016). Lateral hypothalamic circuits for feeding and reward. Nature neuroscience, 19(2), 198-205. Chicago
\\
\\
$^{54}$ Alhadeff, A. L., Hayes, M. R., \& Grill, H. J. (2014). Leptin receptor signaling in the lateral parabrachial nucleus contributes to the control of food intake. American Journal of Physiology-Regulatory, Integrative and Comparative Physiology, 307(11), R1338-R1344.
\\
\\
$^{55}$ Olsen, S. R., \& Wilson, R. I. (2008). Cracking neural circuits in a tiny brain: new approaches for understanding the neural circuitry of Drosophila. Trends in neurosciences, 31(10), 512-520.
Chicago
\\
\\
$^{56}$ De Velasco, B., Erclik, T., Shy, D., Sclafani, J., Lipshitz, H., McInnes, R., \& Hartenstein, V. (2007). Specification and development of the pars intercerebralis and pars lateralis, neuroendocrine command centers in the Drosophila brain. Developmental biology, 302(1), 309-323.
\\
\\
$^{57}$ Pool, A. H., Kvello, P., Mann, K., Cheung, S. K., Gordon, M. D., Wang, L., \& Scott, K. (2014). Four GABAergic interneurons impose feeding restraint in Drosophila. Neuron, 83(1), 164-177.
\\
\\
$^{58}$ Cooper GM. (2000). The Cell: A Molecular Approach (Vol. 2). Sunderland (MA): Sinauer Associates.
\\
\\
$^{59}$ Harwood, C., \& Wipat, A. (2013). Microbial Synthetic Biology (Vol. 40). Elsevier.
\\
\\
$^{60}$ Hobert, O. (2011). Regulation of terminal differentiation programs in the nervous system. Annual review of cell and developmental biology, 27, 681-696.
\\
\\
$^{61}$ Fraser, A. G., Kamath, R. S., Zipperlen, P., Martinez-Campos, M., Sohrmann, M., \& Ahringer, J. (2000). Functional genomic analysis of C. elegans chromosome I by systematic RNA interference. Nature, 408(6810), 325.
\\
\\
$^{62}$ Ng, W. L., \& Bassler, B. L. (2009). Bacterial quorum-sensing network architectures. Annual review of genetics, 43, 197-222.
\\
\\
$^{63}$ Wingard, J. N., Chan, J., Bosanac, I., Haeseleer, F., Palczewski, K., Ikura, M., \& Ames, J. B. (2005). Structural analysis of Mg$^{2+}$ and Ca$^{2+}$ binding to CaBP1, a neuron-specific regulator of calcium channels. Journal of Biological Chemistry, 280(45), 37461-37470.

\end{document}